\documentstyle[12pt,A4,psfig]{article}

\newcommand{\be}{\begin{equation}}
\newcommand{\ee}{\end{equation}}

\def\journalfont{\it}      % this allows redefinition of the font later
\def\jou#1{{\journalfont #1\ }}
\def\pl{\jou{   Phys.\ Lett.}}
\def\pr{\jou{   Phys.\ Rev.}}

\def\prl{\jou{  Phys.\ Rev.\ Lett.}}
\def\np{\jou{  Nucl.\ Phys.}}
\def\apj{\jou{  Astrophys. J.}} 
\def\ut#1{\mathop{\vtop{\ialign{##\crcr
     $\hfil\displaystyle{#1}\hfil$\crcr\noalign
     {\kern1pt\nointerlineskip}\hbox{$\hfil\sim\hfil$}\crcr
     \noalign{\kern1pt}}}}}

\begin{document}
 \begin{titlepage}
 \parindent 0pt
 \font\big=cmbx10 scaled\magstep2
%astro-ph/yymmnn
%Uppsala-TF 02 \hfill\break
\vskip 2cm
%February '96,      \hfill\break
\begin{center}
{\big REVISITING NUCLEOSYNTHESIS CONSTRAINTS ON 
PRIMORDIAL MAGNETIC FIELDS\\}
\vskip 1cm
Dario~Grasso$^{\dag}$ and Hector R.~Rubinstein\\
 \vskip .3cm
{\em Department of Theoretical Physics, Uppsala University\\
Box 803 S-751 08 Uppsala, Sweden, and\\
Department of Physics, University of Stockholm\\
Vanadisv\"agen 9, S-113 46 Stockholm, Sweden.}
\end{center}
\vskip .5cm
\begin{abstract}
In view of several conflicting results, we reanalyze the effects 
of magnetic fields 
on the primordial nucleosynthesis. In the case the magnetic 
field is homogeneous over a horizon volume,
we show that the main effects of the magnetic field are 
given by the contribution  of its energy density to the 
Universe expansion rate and the effect of the field on the 
electrons quantum statistics. 
Although, in order to get an upper limit on the field strength,
the weight of the former effect is numerically larger, 
the latter cannot be neglected. 
Including both effects in the PN code we get the upper limit 
$B \le 1\times 10^{11}$ Gauss at the temperature $T = 10^9~^oK$. 
We generalize the considerations to cases when instead
 the magnetic is inhomogeneous on the horizon length.
We show that in these cases only the effect of the magnetic 
field on the electrons statistics is relevant. 
If the coherence length of the magnetic field at the end of the
PN is in the range $10 \ll L_0 \ll 10^{11}$ cm
our upper limit is $B \le 1\times 10^{12}$ Gauss. 
\end{abstract} 
\vskip 2cm 
$^{\dag}$ EEC Twinning Fellow
 \end{titlepage}
\newpage
\font\big=cmbx10 scaled\magstep1
\font\vbig=cmbx10 scaled\magstep2
\vskip .2cm
\noindent

The study of the effects of  magnetic fields 
on the primordial nucleosynthesis (PN) began with 
the pioneering works of Greenstein \cite{Green}
and Matese and O'Connell \cite{Matese}. 
In a previous paper Matese and O'Connell 
\cite{Matese2} first observed that magnetic fields
larger than the critical value $B_c \equiv eB/m_e^2 = 4.4
\times 10^{13}$ Gauss can affect the $\beta$-decay rate, mainly 
through the modification of 
the phase space of the electrons and positrons.  
Greenstein advanced that, if magnetic fields were present in the 
early Universe, then  this can affect  not only   
the weak processes regulating the neutron to proton ratio in 
the early Universe, namely 
$$ n \,\, + \,\, e^+ \,\, \rightleftharpoons \,\, p\,\, + 
\,\,\overline\nu$$
$$ n \,\, + \,\, \nu \,\, \rightleftharpoons \,\, p\,\, + \,\,e^-$$
$$ n \,\, \rightleftharpoons \,\, p\,\, + \,\,e^- \,\, + \,\,
\overline\nu \ ,  $$
but also the expansion rate of the Universe.
These effects will have consequences on the PN. However,
they are competing effects. In fact, whereas the magnetic field 
energy density $\rho_B = B^2/8\pi$ accelerates the
Universe expansion and therefore increases the predicted 
abundance of the relic $^4{\rm He}$, the effect of the field 
on the weak processes is to increase the $n \rightarrow p$ 
conversion rate reducing the predicted $^4{\rm He}$ abundance.
Greenstein argued that the former effect dominates over the 
latter. In a following paper, Matese and O'Connell \cite{Matese} 
improved Greenstein's considerations by computing explicitly 
$n \rightarrow p$ conversion rate at the PN time in presence 
of a strong magnetic field and they confirmed Greenstein's 
qualitative conclusions.
%\begin{eqnarray}
%\Gamma_{n\,\to\,p}(B) = {\gamma\over \tau}\sum_{n=0}^{\infty} 
%(2 - \delta_{n0})
% \int_{\sqrt{1+2(n+1)\gamma}}^\infty \;d\epsilon
%{\epsilon \over \sqrt{\epsilon^2-1-2(n+1)\gamma}} 
% {1\over 1+e^{{m_e\epsilon\over T}}} \label{rate} \nonumber\\
%\left[{(\epsilon+q)^2 e^{m_e(\epsilon+q)\over T_\nu} 
%\over 1+e^{{m_e(\epsilon + q)
%\over T_\nu}}} + {(\epsilon-q)^2 e^{{m_e\epsilon \over T}}\over
%1+e^{m_e(\epsilon-q)\over T_\nu}}\right]\ .
%\end{eqnarray}
%In the range of temperatures in which PN takes place 
%the effect of the magnetic field is to increase this rate
%almost linearly with the field strength.
The comparison of the PN predictions for the 
relic light isotope abundances with the observations 
provides a criteria to bound the magnetic field strength 
at the nucleosynthesis time \cite{KT}. However, Matese and 
O'Connell did not provide any upper limit.

Recently, the interest about magnetic fields in the early
Universe received new impulse from the idea that these 
fields might be produced during the electroweak phase 
transition\cite{Vacha} or inflation \cite{inflation}. 
The renewed interest in this subject led several authors 
to investigate Matese and O'Connell's treatment again. 

Cheng, Schramm and Truran \cite{Schramm} first used the standard
nucleosynthesis code \cite{code} to get an upper limit on the
magnetic field strength. The effects of the field they 
considered are the same studied by Matese and O'Connell.
However Cheng et al. disagree with Matese and O'Connell in their 
conclusions. In fact, they claim that the effect of the
magnetic field on the weak reactions is the most important.
%Nevertheless, as we shall discuss below, this conclusion looks 
%to be in contrast with their numerical results. 
Cheng at al.  obtained the following upper limit: $B < 10^{11}$
Gauss at the end of the nucleosynthesis ($T = 10^9\  ^oK$).
In both the Matese and O'Connell
and Cheng at al. papers the magnetic field was assumed to be 
uniform over an horizon volume. We will return below to the 
crucial implications of this assumption.

Recently, we reconsidered this subject \cite{us}. Besides the 
effects of the magnetic field on the reaction 
rates, we considered the effect of the field on the 
statistical distribution of the electrons and positrons.
In fact, due to the effect of the magnetic field on the electron 
wave function, the phase space of the lowest Landau level is 
enhanced. If the temperature is small with respect to the energy
gap between the lowest and first exited level, that is if
 $eB \ll T^2$ \footnote{Observe that, neglecting dissipation,
$eB/T^2$ remains constant during the Universe expansion.}, 
a relevant fraction of electron-positron pairs
will {\it condense} in the lowest Landau level affecting 
the statistical distribution of the electron gas. 
As a consequence, the number density and energy density of 
electrons and positrons are increased with respect to the 
case where the magnetic field is not present. The expressions
for the electron+positron energy density and pressure are 
\begin{eqnarray}
\rho_e &=& {\gamma\over 2\pi^2} m_e^4\sum_{n = 0}^{\infty} 
(2 - \delta_{n0}) \int_{\sqrt{1+2(n+1)\gamma}}^\infty \;d\epsilon
{\epsilon^2 \over \sqrt{\epsilon ^2-1-2(n+1)\gamma}} \label{rhoe} \\
& & \left({1\over 1+e^{{m_e\epsilon\over T} + \mu}} 
+ {1\over 1+e^{{m_e\epsilon\over T} - \mu}}\right)
\nonumber \\ 
p_e &=& {\gamma\over 6\pi^2} m_e^4\sum_{n = 0}^{\infty} 
(2 - \delta_{n0}) 
\int_{\sqrt{1+2(n+1)\gamma}}^\infty \;d\epsilon
{(\epsilon^2 - 1) \over \sqrt{\epsilon ^2-1-2(n+1)\gamma}} 
\label{pe} \\
& & \left({1\over 1+e^{{m_e\epsilon\over T} + \mu}} 
+ {1\over 1+e^{{m_e\epsilon\over T} - \mu}}\right)
\nonumber~. 
\end{eqnarray}
Here $\gamma \equiv B/B_c$, $\epsilon$ and $\mu$ are the electron,
or positron, energy and chemical potential in mass units, and 
$n$ labels the Landau level.   

It is evident that the modified contribution of electrons and 
positrons to the total energy density will affect the expansion 
rate of the Universe. The entropy content
of the Universe will be also changed by the effect on the field
on the electrons thermodynamics.
However, for values of the ratio $eB/T^2 \sim 1$, the main
effect of the modified electron statistics on PN is on 
the variation that it induces on the time derivative of the 
photon temperature. This derivative is given by 
\begin{equation}
{dT\over dt} = - 3H {\rho_{em} + p_{em}\over 
d\rho_{em}/dT}
\end{equation}
where $\rho_{em} = \rho_e + \rho_\gamma$, $p_{em} = p_e +
p_\gamma$, $H$ is the Universe expansion rate and $T$ is the
photon temperature. For small values
of the ratio $eB/T^2$, the most relevant
effect of the magnetic field enters in the derivative 
$d\rho_{em}/dT_\gamma$ that is smaller than the value it would
have if the field were not present. 
More physically, this effect can be interpreted as a delay in 
the electron-positron annihilation time induced by the 
magnetic field. This will give rise to a slower entropy transfer
from the electron-positron pairs to the photons, then to a slower
reheating of the heat bath. In fact, due to the enlarged
phase-space of the lowest Landau level of electrons and 
positrons, the equilibrium of the process 
$e^+e^- \leftrightarrow \gamma$ is shifted towards its left side.
 As we already showed in ref.\cite{us},
and we are going to discuss in more details here,
this effect has a clear signature on the Deuterium and $^3$He 
predicted abundances besides that on the $^4$He abundance.

Including the effect of the magnetic field on the
electron statistics, together
with the effect of the field energy density and the effect
on the weak processes in the standard nucleosynthesis 
code \footnote{However we used here an old version of the code,
see \cite{code}.}
we got the upper limit $B \le 3 \times 10^{10}$ Gauss.  
 We also considered other kind of 
effects like that of the magnetic field on the nucleon 
\cite{Bander} and electron masses, but these effects proved 
to be subdominant. 

Recently our conclusions have been put in question in a preprint
by Kernan, Starkman and Vachaspati \cite{KSV}. In this preprint,
the effect of the magnetic field on the PN has been parameterized
in terms of an effective number of neutrinos in order to avoid 
the numerical PN computations.  Kernan et al. considered,
at least in principle, all the effects of the magnetic field
that we discussed 
above, including that on the electron statistics. However,
although for opposite reasons, their main conclusion disagree 
both with Cheng et al. and our results. In fact, in Kernan et al.
opinion the most relevant consequence of the magnetic field 
on the PN is the one mediated by the effect of the field energy 
density on the expansion rate of the Universe.  
The upper limit on the strength of the magnetic field they got 
is $eB/T_{\nu}^2 \le 1$ that can be read $B \le 1 \times 10^{11}$
 at $T = 10^9\ ^o{\rm K}$. This result does not differ from 
Cheng at al. \cite{Schramm} numerical result although it does 
in its physical interpretation.  

The aim of this letter is to try to clarify this confused 
situation. As we are going to show, some mistakes are probably 
present in all refs.\cite{Schramm}\cite{KSV}\cite{us}.
To start with, we have recently realized that our upper limit,
in ref.\cite{us}  is incorrect. This mistake was due to 
the use of a wrong version of the standard nucleosynthesis code. 
We have now rerun the right version of the PN code \cite{newcode}
with an improved subroutine that evaluates the effect of
the magnetic field on the electron and positron energy
density, pressure and energy density time derivative. 
We considered also the effects of the magnetic field
on the weak reaction and the effect of the field energy density 
on the Universe expansion rate.  
Including all possible effects in the PN code we get the 
predictions we report in table I.    

\begin{table}[h]
Table I. Here we report the PN predictions for the relic 
light element abundance for several values of the parameter
$\gamma \equiv B/B_C$ given at the temperature
$T =  10^9~^o{\rm K}$. 

\begin{center}
\begin{tabular}{c|c|c|c}
$\gamma(T = 10^9\,^o{\rm K})$ & $^4$He
& (D+$^3$He)/H & $^7$Li/H \\
\hline\hline
$0$ & $0.237$ & $1.04\times 10^{-4}$ & $1.15\times 10^{-10}$\cr
$1\times 10^{-3}$ & $0.240$ & $1.02\times 10^{-4}$ & $1.17\times
 10^{-10}$\cr
$2\times 10^{-3}$ & $0.242$ & $1.02\times 10^{-4}$ & $1.19\times
 10^{-10}$\cr
$3\times 10^{-3}$ & $0.244$ & $1.03\times 10^{-4}$ & $1.19\times
 10^{-10}$\cr
$4\times 10^{-3}$ & $0.246$ & $1.05\times 10^{-4}$ & $1.20\times 
10^{-10}$\cr
$5\times 10^{-3}$ & $0.249$ & $1.08\times 10^{-4}$ & $1.21\times 
10^{-10}$\cr
$6\times 10^{-3}$ & $0.252$ & $1.12\times 10^{-4}$ & $1.22\times 
10^{-10}$\cr
$7\times 10^{-3}$ & $0.255$ & $1.16\times 10^{-4}$ & $1.24\times 
10^{-10}$\cr
$8\times 10^{-3}$ & $0.258$ & $1.22\times 10^{-4}$ & $1.27\times 
10^{-10}$\cr
$9\times 10^{-3}$ & $0.262$ & $1.28\times 10^{-4}$ & $1.31\times 
10^{-10}$\cr
$1\times 10^{-2}$ & $0.266$ & $1.35\times 10^{-4}$ & $1.36\times 
10^{-10}$\cr
\end{tabular}
\end{center}
\end{table}

These results have been obtained using $N_{\nu} = 3$, the 
lowest value 
of the neutron half life compatible with experimental data,
($\tau_{1/2}(n) = 885\ {\rm s.}$) and the smallest value
of the baryons to photons ratio that makes the predicted 
$({\rm D} + ^3{\rm He})/{\rm H}$ ratio, in absence of the 
magnetic field,
compatible with the astrophysical observations ($\eta = 
2.8 \times 10^{-10}$). This choice assured the minimal 
predicted abundance of the $^4{\rm He}$ for each value
of the magnetic field strength that we considered.
Requiring that such abundance do not exceed the value
of $0.245$ \cite{Olive} we infer from our data the upper 
limit 
\be
B \le 1 \times 10^{11} ~{\rm Gauss}
\ee
when the temperature was $T =  10^{9}\,^o{\rm K}$ (end of PN).  

In order to discriminate the relative weight of the effects
we considered we have run our code leaving on only one effect  
at a time. We start considering only the effect of the 
magnetic field energy density. As expected, we see in 
Tab. II that the predicted $^4$He abundance grows with
the field strength. In fact, larger values of $B$
induce a larger Universe expansion rate then a 
higher freeze-out temperature for the $n/p$ ratio.
  
\begin{table}[h]
Table II. Here we report the PN predictions obtained considering
only the effect of the magnetic field energy density on the
Universe expansion.  

\begin{center}
\begin{tabular}{c|c|c|c}
$\gamma(T = 10^9\,^o{\rm K})$ & $^4$He
& (D+$^3$He)/H & $^7$Li/H \\
\hline\hline
$0$ & $0.237$ & $1.04\times 10^{-4}$ & $1.15\times 10^{-10}$\cr
$1\times 10^{-3}$ & $0.237$ & $1.05\times 10^{-4}$ & $1.15\times
 10^{-10}$\cr
$2\times 10^{-3}$ & $0.238$ & $1.06\times 10^{-4}$ & $1.15\times
 10^{-10}$\cr
$3\times 10^{-3}$ & $0.240$ & $1.09\times 10^{-4}$ & $1.15\times
 10^{-10}$\cr
$4\times 10^{-3}$ & $0.241$ & $1.12\times 10^{-4}$ & $1.16\times 
10^{-10}$\cr
$5\times 10^{-3}$ & $0.244$ & $1.17\times 10^{-4}$ & $1.17\times 
10^{-10}$\cr
$6\times 10^{-3}$ & $0.246$ & $1.22\times 10^{-4}$ & $1.19\times 
10^{-10}$\cr
$7\times 10^{-3}$ & $0.249$ & $1.28\times 10^{-4}$ & $1.23\times 
10^{-10}$\cr
$8\times 10^{-3}$ & $0.252$ & $1.36\times 10^{-4}$ & $1.28\times 
10^{-10}$\cr
$9\times 10^{-3}$ & $0.256$ & $1.44\times 10^{-4}$ & $1.34\times 
10^{-10}$\cr
$1\times 10^{-2}$ & $0.259$ & $1.53\times 10^{-4}$ & $1.42\times 
10^{-10}$\cr
\end{tabular}
\end{center}
\end{table}

In Tab. III we report instead our predictions in the case
only the effect of the magnetic field on the electron quantum 
statistic is considered. 
Note that, in order to display the relative weight of the effects
that we considered, deriving the results of Tab. II and III we
kept the values of $\tau_{1/2}(n)$ and $\eta$ fixed and equal
to the value we used for Tab. I. 

\begin{table}[h]
Table III.  Here we report the PN predictions obtained 
considering only the effect of the magnetic field energy 
density on the electrons and positrons quantum statistics.

\begin{center}
\begin{tabular}{c|c|c|c}
$\gamma(T = 10^9\,^o{\rm K})$ & $^4$He
& (D+$^3$He)/H & $^7$Li/H \\
\hline\hline
$0$ & $0.237$ & $1.04\times 10^{-4}$ & $1.15\times 10^{-10}$\cr
$1\times 10^{-3}$ & $0.240$ & $1.01\times 10^{-4}$ & $1.17\times
 10^{-10}$\cr
$2\times 10^{-3}$ & $0.241$ & $9.99\times 10^{-5}$ & $1.19\times
 10^{-10}$\cr
$3\times 10^{-3}$ & $0.241$ & $9.89\times 10^{-5}$ & $1.20\times
 10^{-10}$\cr
$4\times 10^{-3}$ & $0.242$ & $9.81\times 10^{-5}$ & $1.21\times 
10^{-10}$\cr
$5\times 10^{-3}$ & $0.243$ & $9.72\times 10^{-5}$ & $1.22\times 
10^{-10}$\cr
$6\times 10^{-3}$ & $0.243$ & $9.65\times 10^{-5}$ & $1.22\times 
10^{-10}$\cr
$7\times 10^{-3}$ & $0.243$ & $9.59\times 10^{-5}$ & $1.23\times 
10^{-10}$\cr
$8\times 10^{-3}$ & $0.244$ & $9.51\times 10^{-5}$ & $1.24\times 
10^{-10}$\cr
$9\times 10^{-3}$ & $0.244$ & $9.46\times 10^{-5}$ & $1.24\times 
10^{-10}$\cr
$1\times 10^{-2}$ & $0.244$ & $9.40\times 10^{-5}$ & $1.25\times 
10^{-10}$\cr
$1.1\times 10^{-2}$ & $0.245$ & $9.35\times 10^{-5}$ & $1.26
\times 10^{-10}$\cr
$1.2\times 10^{-2}$ & $0.245$ & $9.28\times 10^{-5}$ & $1.26
\times 10^{-10}$\cr
\end{tabular}
\end{center}
\end{table}

It is interesting to observe the behavior of
the sum of the predicted abundances for the Deuterium and 
$^3$He while varying the magnetic field strength.
This is qualitatively different from what reported
in Tab. II. In fact, if only the effect on the 
electron statistics is considered,  
the (D + $^3$He)/H abundances ratio decreases  
when the field strength increases (see also Fig.2). 
At the same time, 
the predicted $^4$He increases increasing $B$ (see Fig.1).

From the results we reported in Tab. II it follows that
the upper limit we would get if only the effect of the magnetic 
field energy density were present is $B \le 2\times 10^{11}$
Gauss whereas considering only the effect on the electrons 
quantum statistics (see Tab. III)
we would get $B \le 5\times 10^{11}$
\footnote{We have to point out that this limit has been 
obtained using only our predictions for the $^4$He
abundance keeping the value of $\eta$ fixed. However, since
$\eta$ is an unknown parameter, it should be rescaled for
every chosen value of $B$ in such a way to get the largest
D + $^3$He abundance compatible with the observational data
(see below).}.  
Thus, contrary to our previous claim \cite{us},
 the former effect numerically dominates the latter.
However, from Fig. 1 it is clear at a glance that the effect
of the magnetic field on the electron statistics cannot be 
neglected and indeed it dominates for small values of the
field strength.   
In this sense we still disagree with the conclusion of the 
authors of ref.\cite{KSV}. 
It may be that in ref.\cite{KSV} the effect of the magnetic 
field on the electrons and positrons statistical distribution 
was not treated properly close to the saturation of the upper
limit on B, that is for  $eB/T^2 \sim  1$. 

Concerning the effect of the magnetic field on the weak reactions
we have verified that such effect is negligible 
with respect to the others. Mass changes are also negligible
in the range of field strengths that we considered.
 About these conclusions we agree with the
authors of the refs.\cite{Matese} and \cite{KSV}.
Since the numerical predictions of ref.\cite{Schramm}
do not differ significantly from those of ref.\cite{KSV}
and the results we reported in Tab. II we think 
that the different physical interpretation of the
results given in ref.\cite{Schramm} is probably due to
an oversight writing that paper.

To summarize, we have found that the main effects of
a magnetic field on the PN plays through the action of
the field on the electrons and positrons quantum statistics
and the direct effect of the field energy density on the
expansion rate of the Universe. Although the latter effect
is numerically larger than the former, these effects are indeed
comparable, at least if the magnetic 
field is uniform over the horizon scale as we assumed so far.

 Let us now discuss how our previous consideration will
be affected if we relax the uniformity assumption. 
 Before discussing more model dependent motivations 
to consider inhomogeneous magnetic fields, it is mandatory
to compare the upper limit we got above with the upper limit
one would get considering the effect of an homogeneous cosmic 
magnetic field on the early Universe isotropy. In fact, 
since magnetic fields break rotational invariance they can 
induce an
anisotropy in the Universe expansion. This anisotropy  
can have observable consequences on the isotropy of the cosmic
background radiation and on the PN \cite{ZRS}\cite{Thorne}.
Using these criteria Zeldovich and Novikov \cite{ZRS} got the 
limit $B \le 4 \times 10^{10}$ Gauss for a magnetic field 
homogeneous at least over a horizon volume   
 considered at the temperature $T = 10^9$ K. 
This limit is roughly one order of magnitude more stringent 
than the one we got above as those obtained in 
refs.\cite{Schramm} and \cite{KSV}as well.

Nevertheless, if the magnetic field is inhomogeneous with a 
coherence length much smaller than the Hubble radius we do not 
have to care about the Zeldovich and Novikov upper limit. 
In our opinion this is the most plausible physical scenario.  
It is in agreement with the predictions of most of the
models for magnetic field generation in the early Universe 
\cite{Kronberg}. Among these models we are mainly interested 
in those that predict that magnetic fields are
generated during the electroweak phase transition.
The predicted coherence length of the field at the electroweak
phase transition 
time is $L_0 \approx m_W^{-1} \sim 10^{-16} H_{ew}^{-1}$ (where 
$H_{ew}^{-1}$ is the Hubble radius at that time) 
in the case the transition is second order \cite{Vacha}
or, perhaps a more probable scenario, 
$L_0 \sim R_{bubble} \approx 
10^{-(2 \div 3)}  H_{ew}^{-1}$ for a first order phase 
transition \cite{Kibble}\cite{BBM}. Assuming the magnetic 
field strength decreases only due the Universe expansion
the ratio $L_0/H^{-1}$ remains constant in time. In
both cases it is evident that $L_0 \ll H^{-1}$.
       
Having magnetic fields that fluctuate in space on a typical
scale $L_0 \ll H^{-1}$ forces us  to reconsider the analysis we 
made in the first part of this letter paying some attention 
to the different scales on which the effects of the field on the
PN act and the nature of the averaging processes.

\begin{figure}[t]  % produce figure here
%\vspace{9pt}
\psfig{file=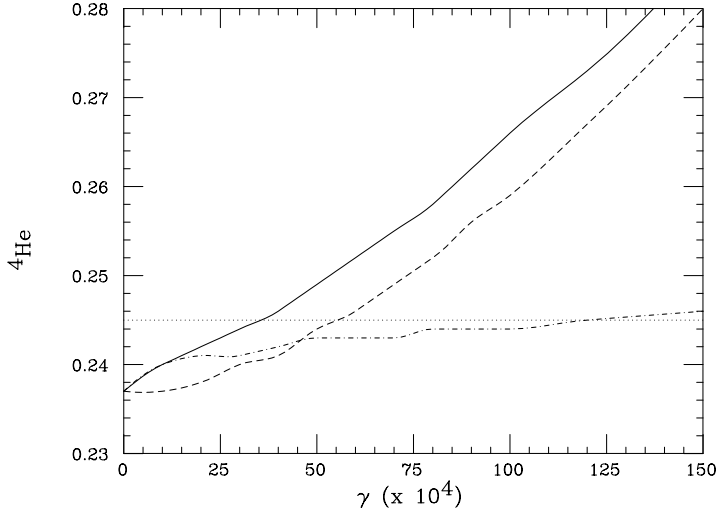,width=11cm}
%\centerline{\hbox{\psfig{figure=he4.ps,width=11cm}}}  
%\label{fig:Fig.1}
\caption{ The $^4$He predicted abundance is represented
in function of the parameter $\gamma$, considered at 
$T = 10^9~^oK$, in three different cases: 
only the effect of the magnetic field energy density 
is considered (dashed line); only the effect of
the field on the electron statistics is considered (dotted-dashed
line); both effects are considered (continuous line).
The dotted line represents the observational upper limit.}
\end{figure}                      

\begin{figure}  % produce figure here
%\vspace{9pt}
\psfig{file=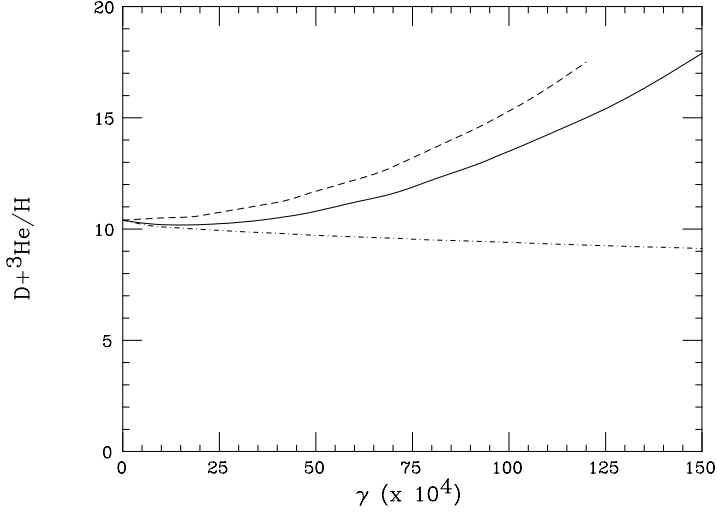,width=11cm}
%\centerline{\hbox{\psfig{figure=deu.ps,width=12cm}}}
%%\label{fig:Fig.2}
\caption{ The (D+$^3$He)/H predicted abundance is represented
for the same cases illustrated in Fig.1.}
\end{figure} 

To start with, let consider the effect of the field on the weak 
processes. The mean free path between two weak reactions is 
$\Gamma_W \sim G_F^2 T^5$ that is of the order of the horizon 
radius at the onset of the PN 
($H^{-1}(T \sim 10^{10}~^oK) \sim 10^{10}$ cm). 
It is clear that if $L_0 \ll \Gamma_w^{-1}$ 
the charged particles involved in the process
will feel a magnetic field changing a number of times between 
the two reactions so that the effect of the field will be 
almost averaged out \cite{Enqvist}. At any rate, 
the effect of the field on the weak reaction rates was already 
subdominant in the case $B$ were assumed to be homogeneous.

More relevant is, instead, the effect of the 
inhomogeneity of the magnetic field on the
action that the magnetic field energy density $\rho_B$ 
plays on the Universe expansion. As discussed in
detail in refs.\cite{Enqvist}\cite{ERS}, since in this case
the field is fluctuating in space, only a {\it mean} magnetic 
field within a horizon volume can be defined as a 
meaningful quantity. This quantity is defined by
\be
{\bar \rho_B} = {1\over 2V_H} \int^{L_H}_{L_0} d^3r B^2_{rms}(r)
\label{rhob}
\ee
where 
\be
B_{rms} = B_0 \left({T\over T_0} \right)^2 
\left({L_0 \over L}\right)^p \label{brms}
\ee
is the root-mean-squared field strength. Here $p$ is a parameter
that depends on the statistical configuration of the magnetic
field. Typical values of $p$ are $1/2,\;1,\;3/2$.  
Inserting Eq. (\ref{brms}) in Eq. (\ref{rhob}) it is evident
that ${\bar \rho_B}$ will suffer a huge suppression if
$L_0 \ll L_H$ so that it will have no chances to affect any
more the PN. The impact of the magnetic field on the electron
quantum statistics is much less 
affected than the others by the inhomogeneity of the field.
In fact, in this case, the characteristic length scale we have  
to compare to $L_0$ is the Compton scattering length of the
electrons $\lambda_C$, since this scattering is responsible 
for the electron thermalization, and the radius of periodic 
motion of
the electrons in the plane normal to the magnetic field vector.
At the PN temperature ($T \sim 10^{9\div 10}~^oK$) the Compton 
cross section does not differ significantly from the Thomson 
cross section. Then we have
\be
\lambda_C = (\sigma_T n_e v)^{-1} \sim 10^{-(2\div 5)}\ {\rm cm}.
\ee 
This has to be compared with the Hubble radius at the same 
temperature that is $\sim 10^{10\div 11}$ cm. 
The radius of the classical orbit of the electron in an
over critical magnetic fields is given by
\be
R_n = \left({n + 1\over \gamma}\right)^{1/2}m_e^{-1}
\ee
where $n$ labels the Landau level.
Since the magnetic field plays its main effect of  on the lowest
Landau level it is clear that the radius we have to care about
is much smaller than $\lambda_C$ and 
far below the expected value of $L_0$ at the PN time.

Since, $\lambda_C \ll \Gamma_W^{-1} \ll H^{-1}$ we see that
the effect of the magnetic field on the electron quantum 
statistics provides the best probe of the field at small
length scales.
If $\lambda_C \ll L_0 \ll H^{-1}$ this is the only effect we 
remain with. 

The PN predictions of Tab. III cannot be used in their 
present form to determine, in this case, the upper limit on 
the magnetic field strength. 
In fact, as we pointed-out above, obtaining Tab. III we kept the 
value of $\eta$ fixed. However, $\eta$ is an unknown parameter 
and its value has to be bound from below in order the predicted 
abundances of D and $^3$He do not exceed the relic abundances 
extrapolated from present observations. 
Since in our case the predicted of abundances of these isotopes 
depend on the field strength,
we have to rescale $\eta$ for every value of $B$ in order 
to get their largest predicted abundances compatible with
the observational upper limit. 
According to ref. \cite{Olive} we assume this limit to be
$({\rm D} + ^3{\rm He})/{\rm H} \le 1.1 \times 10^{-4}$ .
Finally, we have to compare
the predicted abundance of the $^4$He, obtained using such value 
of $\eta$, with the maximal abundance compatible with 
observations $^4{\rm He} \le 0.245$ \cite{Olive}. Using these 
procedure we get the upper limit 
\be
B \le 1\times 10^{12}~ {\rm Gauss}. \label{lim2}
\ee
Again, this limit refers to the temperature $T = 10^9~^oK$.
The corresponding value of $\eta$ is 2.3.  As expected,
this value is smaller than the standard lower limit 
$\eta \ge 2.8$, since the effect of the magnetic field on 
the electron statistics suppresses the predicted abundance of 
D and $^3$He. It is worthwhile to note here that, although 
this effect increases
also the predicted abundance of $^4$He, large magnetic fields 
might help to alleviate the ``nucleosynthesis crisis" that is 
presently under debate \cite{crisis}.

So far we neglected any dissipative effect in the plasma. 
However, even in a relativistic plasma conductivity $\sigma$ is 
a finite quantity. This means that any field configuration on a 
length scale $L_0 \ll L_{diss}$, where 
\be
L_{diss} = \sqrt{{H^{-1}\over 4\pi\sigma}} \ ,
\ee
will be dissipated. It is interesting to observe that if
$eB/T^2 \gg 1$ the conductivity properties of the plasma
will be seriously affected by the magnetic field. In fact, in 
this case the $e^+e^-$ pair number density is increased by the 
effect of the field on the electrons and positrons phase-space.
As a consequence this will make the plasma conductivity to
decrease then $L_{diss}$ to grow \cite{DG}. This can have
important consequences on the field evolution after it
is generated in the early Universe. However, such effect is 
negligible at the PN time. Following ref. \cite{ERS}
$L_{diss}$ at $T = 10^9~^oK$ MeV can be estimated to be
\be
L_{diss} \simeq 0.1 g_*^{-1/4} \left(10^9\over T\right)^{7/4}
\approx 1\div 10 ~{\rm cm}.
\ee 

Taking this considerations into account we conclude that 
our upper limit (\ref{lim2}) applies only if  
$10 \ll L_0 \ll 10^{11}$ cm. 

It is worthwhile to see how our upper limit
constraints at least one of the recent models of relic magnetic 
field generation. Assuming the field is generated at the end 
of a first order electroweak phase-transition 
\cite{Kibble}\cite{BBM}, within large theoretical uncertainties
the predicted magnetic field strength at the temperature 
$T \sim 10^{15}~^oK$ is $\approx 10^{24}$ Gauss
with a coherence length $L_0 \approx 10^{-(2\div 3)}$ cm.
If the magnetic field flux above this length scale remains frozen
in the plasma, so that the field strength decrease only due to 
the Universe expansion, at the PN time we expect: 
$B(T = 10^9~^oK) \sim 10^{12}$ Gauss and 
$L_0 \sim 10^{4\div 5}$ cm. Thus, this model is not excluded 
by PN considerations. 

In conclusion, in this letter we unfortunaly revise upwards
previous limits on homogeneous magnetic field strengths at PN 
time by a factor 10. Earlier calculations contained mistakes on 
a variety of points as discussed. As a consequence, previous 
bounds based on Universe isotropy remain more stringent\cite{ZRS}.
However, in the case the magnetic field is inhomogeneous on 
the horizon length scale, isotropy considerations do not apply.
In this case, we showed that the main observable effect of
the magnetic field on the PN is on the electron and positron 
quantum statistics.

\section*{Acknowledgments}

The authors wish to thank V. Semikoz for valuable discussions. 
This work was supported in part by the Swedish Research Council 
and a Twinning EEC contract.

\end{document}